\documentclass[prb,showpacs,amsmath,amssymb,twocolumn]{revtex4-1}
\usepackage[usenames,dvipsnames]{color}
\usepackage{graphicx,textcase}
\usepackage{dcolumn,overpic}
\usepackage[normalem]{ulem}

\begin{document}

\title{Microscopic origin of Heisenberg and non-Heisenberg exchange interactions in ferromagnetic bcc Fe}
\author{Y.O. Kvashnin$^1$, R. Cardias$^{2}$, A. Szilva$^1$, I. Di Marco$^1$, M.I. Katsnelson$^{3,4}$, A.I. Lichtenstein$^{4,5}$, L. Nordstr\"om$^1$, A.B. Klautau$^{2}$ and O. Eriksson$^1$}
\affiliation{$^1$ Department of Physics and Astronomy, Division of Materials Theory, Uppsala University, Box 516, SE-75120 Uppsala, Sweden}
\affiliation{$^2$ Faculdade de Fisica, Universidade Federal do Para, Belem, PA, Brazil}
\affiliation{$^3$ Radboud University of Nijmegen, Institute for Molecules and Materials, Heijendaalseweg 135, 6525 AJ Nijmegen, The Netherlands}
\affiliation{$^4$ Theoretical Physics and Applied Mathematics Department, Ural Federal University, Mira Str.19,  620002, Ekaterinburg, Russia}
\affiliation{$^5$ Institute of Theoretical Physics, University of Hamburg, Jungiusstrasse 9, 20355 Hamburg, Germany}

\begin{abstract}
By means of first principles calculations we investigate the nature of exchange coupling in ferromagnetic bcc Fe on a microscopic level.
Analyzing the basic electronic structure reveals a drastic difference between the $3d$ orbitals of $E_g$ and $T_{2g}$ symmetries.
The latter ones define the shape of the Fermi surface, while the former ones form weakly-interacting impurity levels.
We demonstrate that, as a result of this, in Fe the $T_{2g}$ orbitals participate in exchange interactions, which are only weakly dependent on the configuration of the spin moments and thus can be classified as Heisenberg-like.
These couplings are shown to be driven by Fermi surface nesting.
In contrast, for the $E_g$ states the Heisenberg picture breaks down, since the corresponding contribution to the exchange interactions is shown to strongly depend on the reference state they are extracted from.
Our analysis of the nearest-neighbour coupling indicates that the interactions among $E_g$ states are mainly proportional to the corresponding hopping integral and thus can be attributed to be of double-exchange origin.
\end{abstract}

\maketitle

Iron is one of the most abundant elements in the universe.
Its elemental phase has several polymorphs and among those the most stable crystal structure at ambient conditions is a body-centered cubic (bcc) one. 
Ferromagnetism is known to play a decisive role in defining the stability of this structure.\cite{Zener-Fe,Hasegawa-Fe-stab}
The bcc phase is ferromagnetic (FM) up to critical temperature ($T_c$) of 1045 K. 
Quite importantly, above the Curie point, the bcc structure is preserved in a certain temperature range before it undergoes a transition to the fcc phase. 
This fact implies that the local magnetic moments exist in the paramagnetic (PM) phase, where strong short-ranged magnetic order was proposed.\cite{Fe-short-range}
The magnetism of Fe is of mixed itinerant and localised nature and it is a matter of debate, which model describes it the best.
Both the $T_c$ and the magnetic excitation spectra at low temperatures can be well described by means of the Heisenberg Hamiltonian (HH), parameterised by \textit{ab initio} calculations.\cite{exch1,lichtenstein,halilov,antropov-99,exch2,pajda,rspt-jijs}
However, in several works\cite{attila-prl,Fe-biquad} it was argued that in order to describe a large palette of magnetic states, higher-order (biquadratic) exchange interactions have to be taken into account.
The results of the self-consistent spin spiral calculations also indicate that the magnitude of the magnetic moment in bcc Fe can differ by almost 30$\%$ in various states,\cite{Ruban-spirals,Okatov-magnetoelastic} which in principle disagrees with the assumptions of the Heisenberg picture.
Indeed, it was previously pointed out that the parameterisation of HH for bcc Fe depends on the magnetic configuration they are extracted from.\cite{turzhevskii,Heine-FM-DLM,Bottcher}

The strong correlation effects are known to be important for bcc Fe at finite temperatures, as was shown in Ref.~\onlinecite{AL-PRL01} by means of density functional theory plus dynamical mean field theory (DFT+DMFT) calculations.
Earlier, it was suggested from a qualitative analysis of the electronic structure that the $E_g$ electrons in iron are much more correlated than the $T_2g$ ones.\cite{Fe-FermiLiquid}
This statement was quantitatively checked and partially confirmed in Ref.~\onlinecite{katanin2010}, where the authors performed DFT+DMFT calculations for PM phase of bcc Fe.
According to Ref.~\onlinecite{katanin2010}, the $E_g$ and $T_{2g}$ states have to be analysed separately in this system.
Both types of orbitals were found to contribute equally to the formation of the local moment in its PM phase.
Recently Igoshev~\textit{et al.}\cite{katanin2015} has presented an analysis of the orbital-resolved dynamical susceptibility again using DFT+DMFT formalism. 
The $E_g-T_{2g}$ exchange interactions were suggested to play the main role in the magnetic couplings.

In this work we perform an orbital-by-orbital analysis on the magnetic interactions in FM bcc phase and make an attempt to classify them and associate with the well-known textbook exchange mechanisms.
Surprisingly we find that there is a strong antiferromagnetic (AFM) component to the nearest neighbour (NN) exchange interaction for the states of $T_{2g}$ character.
This is caused by the Ruderman-Kittel-Kasuya-Yosida (RKKY)\cite{rkky}-type coupling, governed by the topology of the Fermi surface (FS). 
In contrast, the $E_g$ states contribute ferromagnetically to the NN coupling with a combination of double exchange (DE) and super-exchange.  
As a consequence, the $E_g$ states give rise to short ranged magnetic interactions in bcc Fe, whereas $T_{2g}$ states contribute to longer range couplings, with a pronounced oscillatory behaviour.

All calculations were performed with the use of standard DFT technique by means of either real-space linear muffin-tin orbital method within the atomic sphere approximation (RS-LMTO-ASA)\cite{rs-lmto1,rs-lmto2} or a full-potential realisation of the LMTO method\cite{rspt-book}.
In this work we concentrate on the ferromagnetic phase, where the many-body effects are known to be suppressed and do not significantly influence the effective magnetic interactions.\cite{exch2,rspt-jijs}
Thus we neglected the effects of strong correlations and employed standard local spin density approximation (LSDA) for the exchange-correlation energy.
The inter-site exchange integrals ($J_{ij}$'s) were extracted by means of magnetic force theorem (MFT).\cite{lichtenstein}
Within this approach the magnetic sub-system is mapped onto HH of the following form:
\begin{equation}
\hat H = - \sum_{i \neq j} J_{ij} \bf{e_i} \cdot \bf{e_j} ,
\label{HH}
\end{equation}
where $\bf{e_i}$ denotes the unit vector along the magnetic moment at the site $i$.
For some calculations we have also adopted a recent generalisation of the MFT allowing for treatment of the non-collinear spin structures.\cite{attila-prl}
In addition to the total value of the $J_{ij}$, we have computed the individual orbital contributions to each particular coupling (for details see e.g. Ref.~\onlinecite{korotin-jijs-wannier}).
The latter ones were grouped according to the representations of the cubic space group, so that each exchange integral was represented as:
\begin{equation}
J_{ij}=J_{ij}^{E_g-E_g}+J_{ij}^{E_g-T_{2g}}+J_{ij}^{T_{2g}-T_{2g}},
\label{maindef}
\end{equation} 
where, for instance, $J_{ij}^{E_g-T_{2g}}$ denotes an aggregate strength of the coupling of the $E_g$ orbitals located on the site $i$ interacting with the $T_{2g}$ subset located at the site $j$.
For an arbitrary $i-j$ pair, the aforementioned \textit{mixed} coupling are allowed by symmetry even in the $Im\bar{3}m$ space group.
This is so, because $J_{ij}$ is an inter-site quantity and thus depends on the bonding vector $\bf{R_{ij}}$, which in most cases locally destroys the full cubic symmetry, hence allowing for mixing between the $E_g$ and $T_{2g}$ orbitals.

First, we have performed a set of calculations by rotating a single Fe moment in a FM background.
For each chosen value of the angle ($\theta$) the electronic structure was calculated and then the $J_{ij}$-parameters were extracted following the recipe given in Ref.~\onlinecite{attila-prl}.
The magnitude of the rotated magnetic moment was constrained to be 2.2 $\mu_B$, which corresponds to the value for the background spins.

\begin{figure}[!h]
	\includegraphics[angle=0,width=85mm]{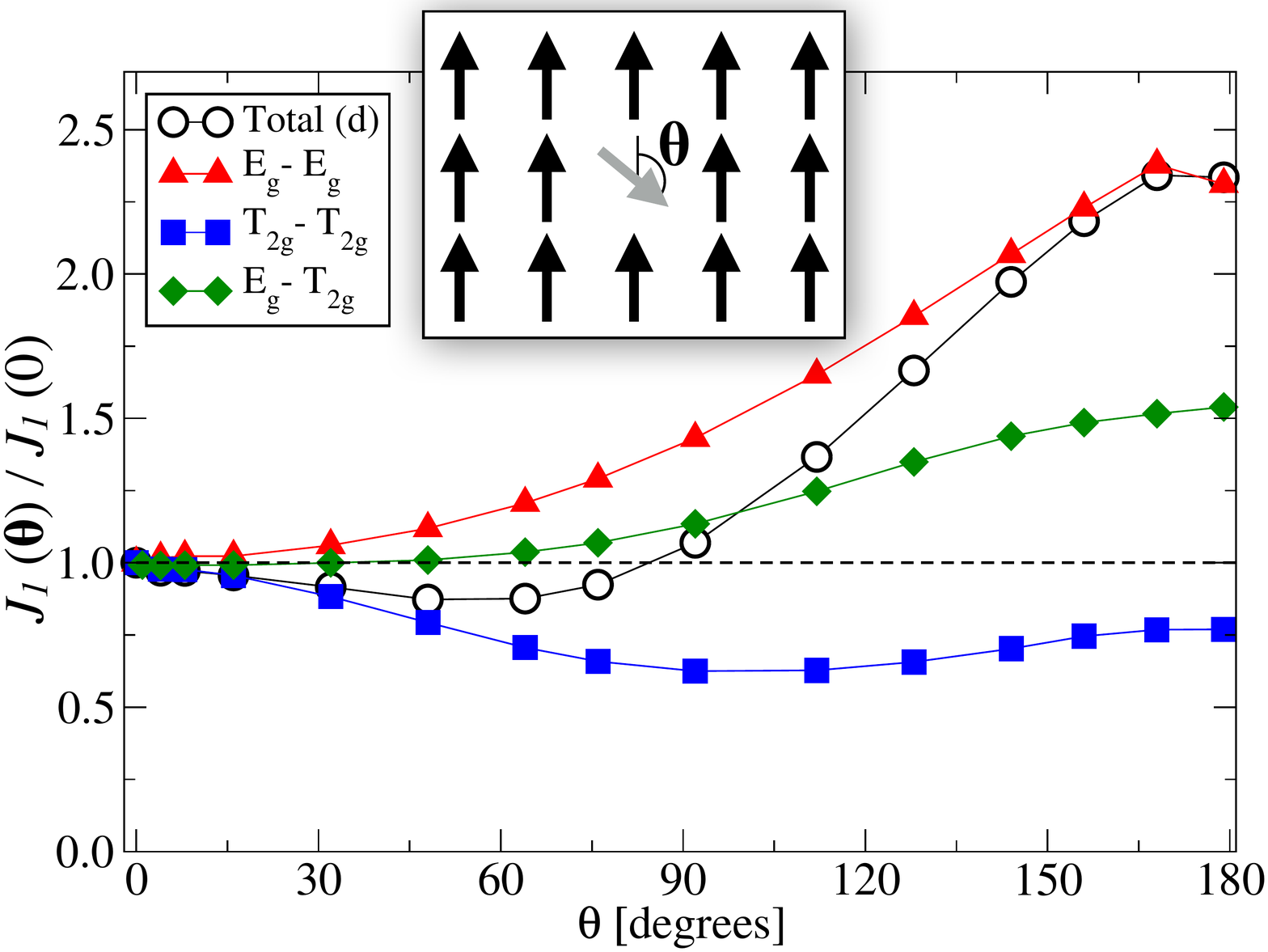}
	\caption{Relative change of the NN exchange interaction in bcc Fe as a function of the rotation angle $\theta$ of a single spin in the FM environment. Each component is renormalised separately. Inset: Schematic picture of the computational setup.}
	\label{J1-vs-theta}
\end{figure}

Fig.~\ref{J1-vs-theta} shows the numerical values of the relative changes of the NN exchange parameter ($J_1$) as a function of the rotation angle $\theta$ in bcc Fe.
Every set of results is normalised with respect to the values for $\theta=0$.
The results for $\theta$ = 0 correspond to the collinear case where all spins are parallel, thus representing the FM reference state.
For FM state we obtained the following values of the contributions to the $J_1$: $J_1^{E_g-E_g}$ = 0.57 mRy, $J_1^{E_g-T_{2g}}$ = 1.37 mRy and $J_1^{T_{2g}-T_{2g}}$ = -0.91 mRy.
One can see that the $E_g-E_g$ and $E_g-T_{2g}$ contributions are FM, while the $T_{2g}-T_{2g}$ one is actually AFM.
Such strong AFM contributions are surprising for bcc Fe, which is known as one of the most stable ferromagnetic materials.
Our results indicate that all three symmetry-resolved contributions have comparable strength and therefore compete with each other.

An closer inspection of Fig.~\ref{J1-vs-theta} leads to an interesting observation. 
Approaching the value of $\theta$=180$^{o}$, the total value of $J_1$ is more than double of its initial value, which indicates that the Heisenberg picture, where the exchange constant is supposed to be independent of the mutual orientation of the spins, breaks down.
However, the analysis of the orbital decomposition reveals that the $\theta$-dependence comes primarily from the $E_g-E_g$ term.
This contribution exhibits the most significant changes and is the main source of the $J_1$ enhancement.
In contrast, the $T_{2g}-T_{2g}$ contribution deviates from its $\theta=0$ value by not more than 37$\%$.
Hence magnetic excitations of these orbitals are better to be described with the HH.
The $E_g-T_{2g}$ contribution shows an intermediate behaviour between the two.

The pronounced $\theta$-dependence of the $J_1^{E_g-E_g}$ might have several reasons.
One has to remember that the exchange constants, computed using MFT are \textit{effective} parameters, and are obtained by imposing a certain shape of the spin Hamiltonian (given by Eq.~(\ref{HH})). 
Thus, all non-Heisenberg terms, influence them.
The most plausible candidates for these terms are either higher order exchange interactions (e.g. biquadratic ones) or DE.
The latter one is quite peculiar, since it also can not be described in terms of Eq.~(\ref{HH}).\cite{anderson-DE,auslender-DE}

\begin{figure}[!h]
	\includegraphics[angle=0,width=70mm]{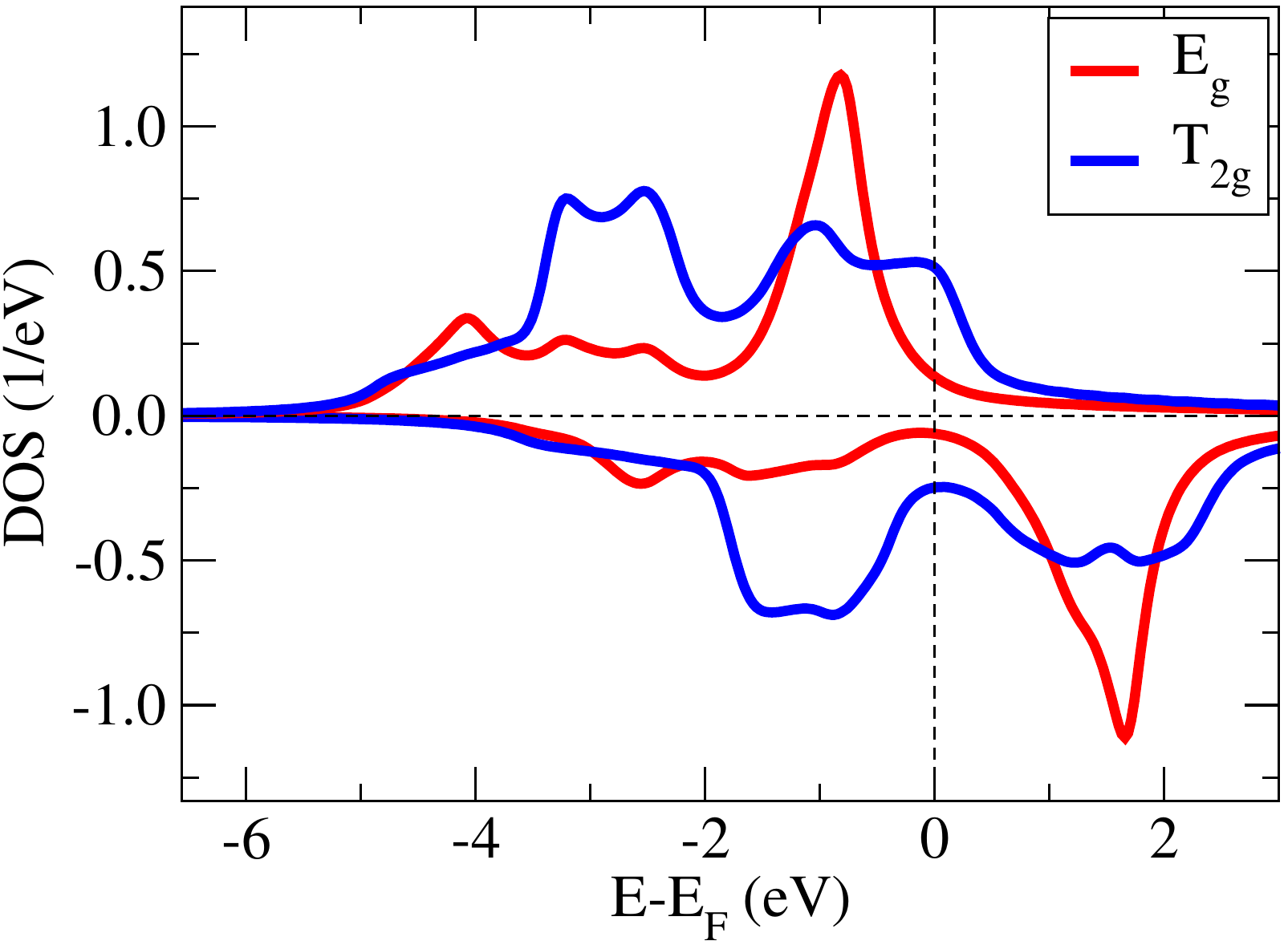}
	\caption{Projected DOS in bcc Fe. Density of the minority states is shown as negative. The spectral function $N(\varepsilon)=-\frac{1}{\pi} \Im G^{\sigma}_{mm} (\varepsilon + i\delta)$ is plotted for $\delta$=0.01 Ry, which produces an artificial Lorenzian broadening of its features.}
	\label{PDOS}
\end{figure}

As was shown above, the $J_1$ coupling in bcc Fe consists of FM and AFM contributions, which have similar magnitude.
Here we argue that in this particular case it is possible to attribute each contribution to different microscopic mechanisms.
To analyze this we show in Fig.~\ref{PDOS} the orbital-projected density of states (DOS) in bcc Fe.
One can see that the bands originating from the $T_{2g}$ states are much broader than those of the $E_g$ ones.
Such a drastic difference has already been emphasised by Katanin \textit{et al.}\cite{katanin2010}, when they addressed its PM phase.
The authors of Ref.~\onlinecite{katanin2010} proposed that the $E_g$ electrons are influenced by strong correlation effects, which leads to their non-Fermi-liquid behaviour.
Here we mention that Ref.~\onlinecite{katanin2010} concerns the paramagnetic phase, which is $\textit{a priori}$ more correlated than the ordered phase, which we discuss here.
The primary reason for it is that at low temperatures the spin fluctuations almost do not contribute to the scattering of the quasiparticles.

What is more important from the DOS figure is that the FS is almost entirely formed by the $T_{2g}$ states, which was already pointed out in Ref.~\onlinecite{bccFe-ARPES}.
In contrast, as we have just shown, the $E_g$ orbitals form a set of half-filled quasi-impurity states.
Such a clear difference is expected to lead to a very pronounced difference concerning the nature of the exchange couplings.
For instance, mechanisms associated with details of the Fermi surface, like the RKKY interaction, are expected to be more pronounced for the $T_{2g}$ states, since these states dominate the Fermi surface.

\begin{figure}[!h]
	\includegraphics[angle=0,width=80mm]{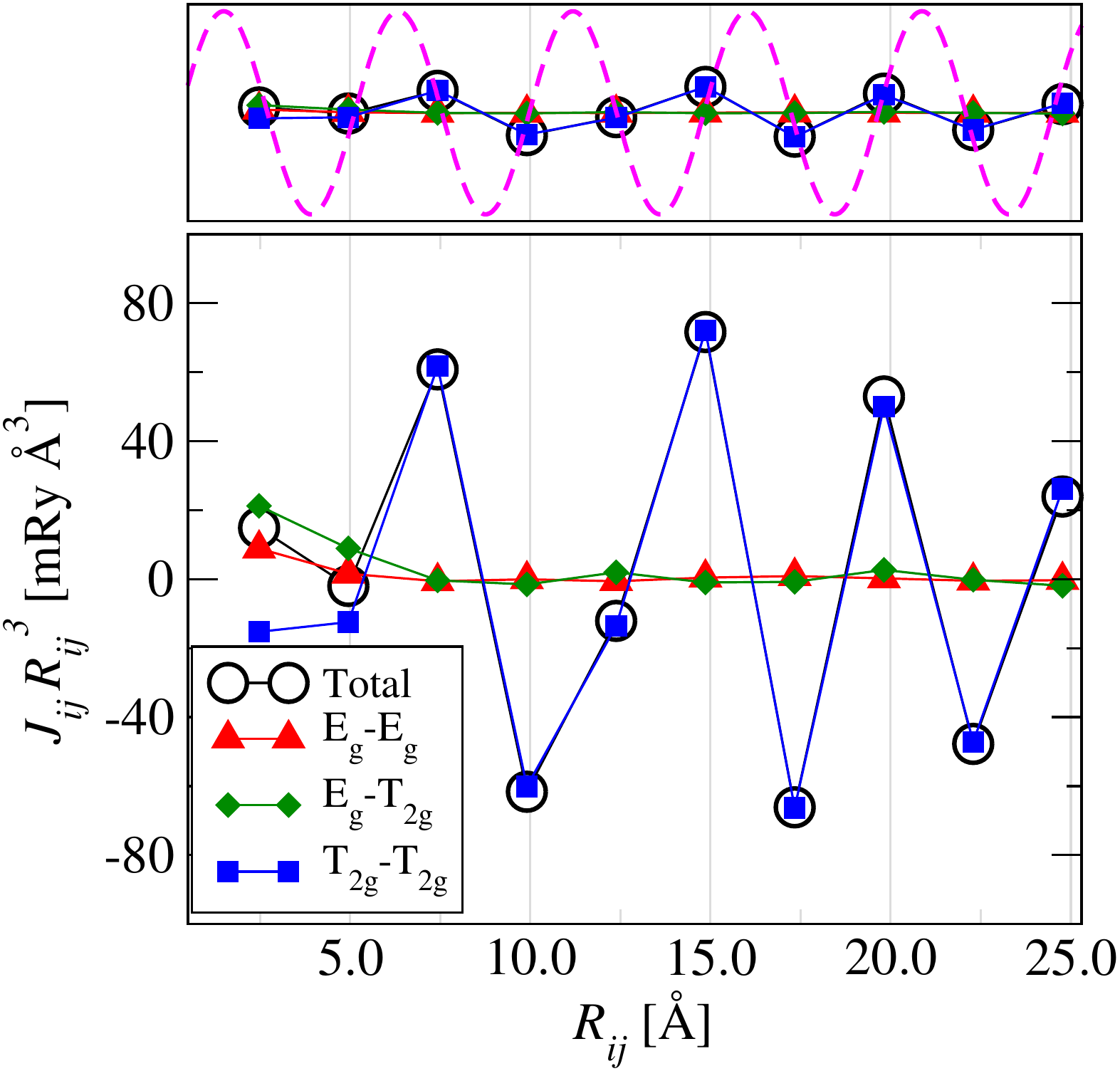}
	\caption{Bottom panel: $J_{ij}R_{ij}^3$ with the neighbours selected to lie along the (111) direction in bcc Fe. $R_{ij}$ is the inter-site distance. 
	Top panel: The same data plotted together with an analytical function $y=A_0\sin{(1.3R_{ij}+\phi_0)}$ (dashed curve), which was obtained from the FS analysis. Only $A_0$ and $\phi_0$ were fitted, see text.}
	\label{rkky}
\end{figure}

To address these issues further, we have analysed the long-ranged magnetic interactions in bcc Fe along several high-symmetry directions.
It was found that $T_{2g}-T_{2g}$ interactions indeed has pronounced RKKY oscillations, in particular along the (111) direction, i.e. the direction along the NN bonding vector.
For this specific direction, we show in Fig.~\ref{rkky} $J_{ij}R_{ij}^3$ as a function of the inter-site distance $R_{ij}$, resolved into different symmetry components.
One can see that the $E_g-E_g$ and $E_g-T_{2g}$ contributions decay rather quickly and are already negligible for the 3rd NN along this path.
The $T_{2g}-T_{2g}$ part, on the contrary, is extremely long-ranged  and gives the main contribution to the total coupling at large distances.
Thus it is clear that the $T_{2g}$ electrons are responsible for a RKKY exchange in bcc Fe.

Furthermore, we analyzed the period of the observed oscillations with emphasis on the features of the FS.
The exchange couplings along this direction are primarily defined by the excitations, carrying the momenta parallel to the high-symmetry line $\Lambda$ (i.e. along $\Gamma$-P direction).
In fact, in this part of the Brillouin Zone the FS topology of both spin channels is extremely simple.
It is characterised by the presence of one electron pocket per spin channel (see e.g. Refs.~\onlinecite{halilov,bccFe-ARPES}). 
Moreover, the orbitally-projected band structures presented in Ref.~\onlinecite{bccFe-ARPES} suggest that these bands have mainly $T_{2g}$ character, which is in line with our conclusions about the origin of the RKKY oscillations.
The fact that the electron pockets for both spin projections are quite isotropic implies that the long-ranged oscillations should possess a single period.
We have analysed in detail the band structure and obtained the following values for the Fermi vectors: ${k}^{\uparrow}_F$= 0.94 \AA$^{-1}$ and ${k}^{\downarrow}_F$= 0.36 \AA$^{-1}$.
The period of the RKKY oscillations is defined by the calliper vector,\cite{pajda} i.e. $k^{\uparrow}_F+k^{\downarrow}_F\approx$1.30 \AA$^{-1}$.
We fitted the computed oscillatory exchange interactions of Fig.~\ref{rkky} with a $sin$-function of a fixed period, as discussed above.
The result is shown in Fig.~\ref{rkky}. 
One can see that the analytical results reproduces excellently the outcomes of our numerical calculations.
This is a strong evidence that the $T_{2g}$ states primarily participate in the Heiseberg-like exchange interactions, driven by RKKY mechanism.

\begin{figure}[!h]
	\includegraphics[angle=0,width=60mm]{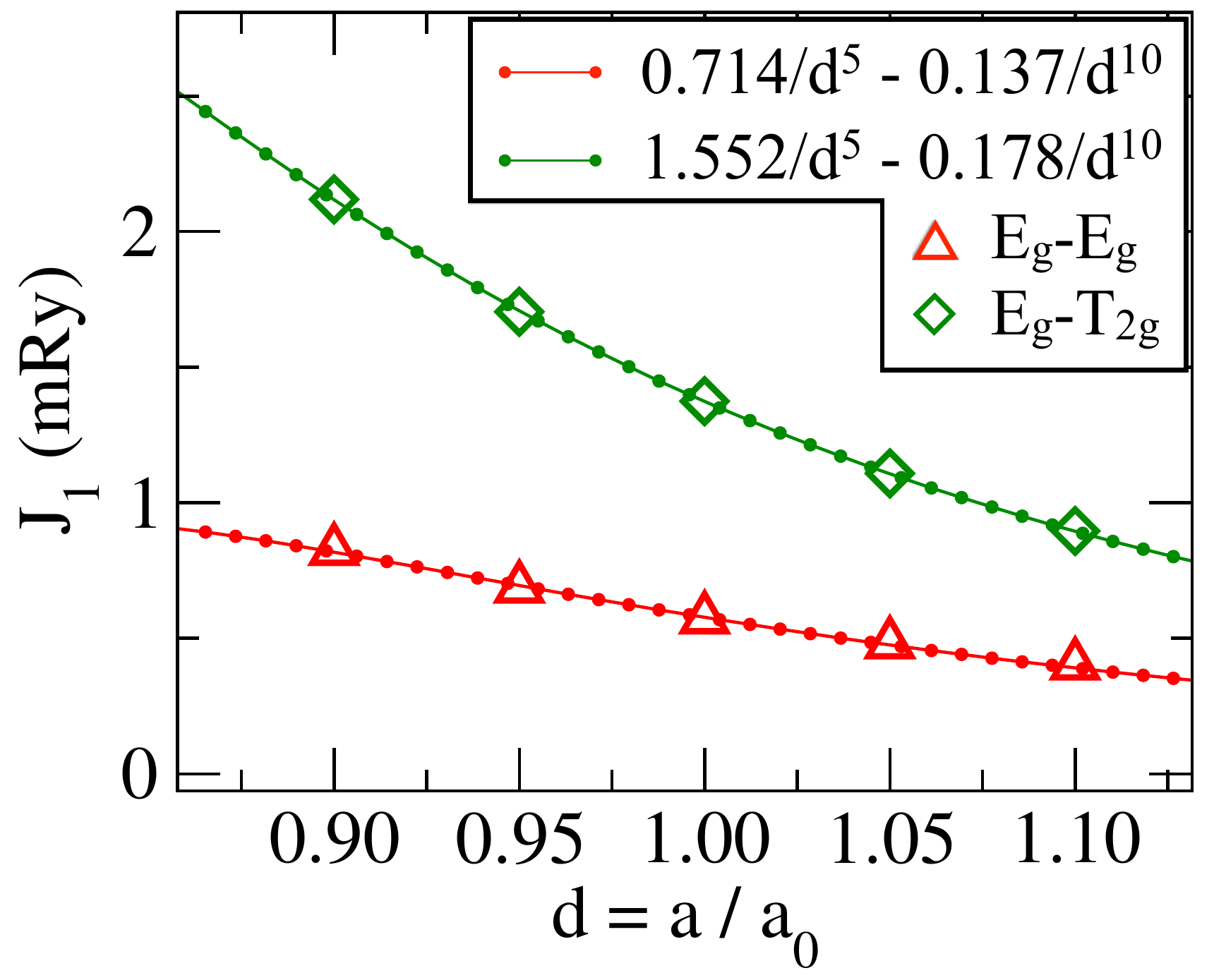}
	\caption{Distance dependence of the $J^{E_g-E_g}_1$ and $J^{E_g-T_{2g}}_1$ interactions in bcc Fe. $a_0$ corresponds to the equilibrium lattice constant (2.86 \AA). To facilitate the analysis, the magnetic moment was constrained to be 2.2 $\mu_B$ at every point.}
	\label{j1dist}
\end{figure}

In contrast, the $E_g$ electrons are involved in other types of magnetic interactions.
In order to shed light on their nature, we have performed an analysis based on the tight-binding picture.
The main quantity in this theory is the inter-site hopping integral ($t$).
The hopping integral between two $3d$ orbitals ($l$=2) located at different sites is expected to scale as $d^{-5}$, where $d$ is the distance between the sites (see e.g. Ref.~\onlinecite{harrison}).
Typically, different contributions to the exchange couplings, scale with different powers of $t$.
For instance, the FM DE is proportional to $t$ ($\propto d^{-5}$), while the AFM super-exchange has a $t^2$-dependence ($\propto d^{-10}$).
In order to identify their relative contributions, we have performed DFT calculations for bcc Fe for different volumes around the equilibrium.
Based on the aforementioned arguments, we performed the fittings of individual orbital contributions to $J_1$ using the following expression:
\begin{eqnarray}
J_1 = \alpha d^{-5} - \beta d^{-10},
%a'=a*V ; b'=b*V^2
\label{de-se}
\end{eqnarray}
where $\alpha$ and $\beta$ are fitting parameters.
The results of the \textit{ab initio} calculations along with the fitted functions are shown in Fig.~\ref{j1dist}.
Both $E_g-E_g$ and $E_g-T_{2g}$ components of the $J_1$ could successfully be fitted.
The obtained values of $\alpha$ and $\beta$ clearly indicate that the FM DE contribution strongly prevails over the AFM part.
Moreover, one can see the crucial role of the $E_g-T_{2g}$ interactions on the ferromagnetism of iron: it provides the dominant FM component of the $J_1$ coupling. 
This observation is in line with a recent study by Igoshev \textit{et al.}\cite{katanin2015} who analysed the paramagnetic susceptibility in bcc Fe and also emphasised an importance of the mixed $E_g-T_{2g}$ interactions for the formation of the local moment. 
The results shown in Fig.~\ref{j1dist} show that the primary contribution to $J_1^{E_g-E_g}$ and $J_1^{E_g-T_{2g}}$ is proportional to the first power of the effective hopping integral $t$.
Moreover, we have shown above that the NN $E_g-E_g$ and $E_g-T_{2g}$ interactions have a pronounced $\theta$-dependence, which implies their non-Heisenberg origin.
This allows us to conclude that these interactions are not primarily dominated by biquadratic interactions.
We draw this conclusion because biquadratic exchange interactions are proportional to higher powers of $t$ (see e.g. Ref.~\onlinecite{Mila-biquad}).
Thus, bringing all the evidences together, we conclude that the DE mechanism is the main source of the NN $E_g-E_g$ and $E_g-T_{2g}$ interactions in bcc Fe.

It is worth mentioning that our results were obtained using DFT, which describes the ground state of the system at zero temperature. 
At high temperatures, especially close to the transition to a PM state, the underlying electronic structure will be significantly modified and previously suggested importance of the biquadratic interactions\cite{attila-prl,Fe-biquad} can not be ruled out.

We find a strong competition between FM and AFM contributions to the NN exchange coupling coming from different $3d$ orbitals in FM bcc Fe.
It is shown numerically that the exchange coupling between the $T_{2g}$ orbitals is relatively independent on the mutual orientation of the spins, thus suggesting their Heisenberg-like nature. 
This conclusion is supported by the analysis of the long-ranged exchange couplings along the NN direction.
The period of RKKY oscillations was shown to be related to the nesting of the FS, that is dominated by contributions from the $T_{2g}$ states.
The $E_g$ electrons, on the contrary, produce relatively short-ranged interactions with a substantial non-Heisenberg behaviour. 
Our analysis demonstrate that the $E_g$ states provide inter-atomic exchange due to a double exchange mechanism.

The peculiarities of the electronic structure of FM bcc Fe is reflected in a unique way on the exchange interactions.
The results obtained in this work demonstrate an importance of non-Heisenberg exchange interactions in iron. The origin of these interactions is identified.
The present work opens a way for construction of more sophisticated and detailed models to describe magnetism of iron.

\begin{acknowledgments}
Y.K. is grateful to P.Bruno (ESRF) for useful discussions.
R. C. and A. B. K. acknowledge financial support from CAPES and CNPq, Brazil.
M.I.K. acknowledges a financial support by European Research Council (ERC) Advanced Grant No. 338957 FEMTO/NANO.
O.E. acknowledges the support from VR, eSSENCE, and the KAW foundation.
The authors acknowledge the computational resources provided by the Swedish National Infrastructure for Computing (SNIC) and Uppsala Multidisciplinary Center for Advanced Computational Science (UPPMAX).
\end{acknowledgments}

\end{document}